\documentclass[]{elsarticle}

\usepackage{lineno,hyperref}

\usepackage{graphicx}
\usepackage{pdfpages}
\usepackage{mathtools}
\usepackage{color}
\usepackage{epsfig}
\usepackage{amssymb}
\usepackage{amsmath}
\usepackage{bm}
\usepackage{float}
\usepackage{setspace}
\usepackage{geometry}
\usepackage{comment}

\journal{Journal of Theoretical Population Biology}

\biboptions{authoryear}

\begin{document}

\begin{frontmatter}

\title{Survival in Branching Cellular Populations}

\author[utk]{Adam S. Bryant}

\author[utk]{Maxim O. Lavrentovich}

\address[utk]{Department of Physics \& Astronomy, University of Tennessee, Knoxville, TN 37966 USA}
\cortext[cor]{\textit{Email address:} {lavrentm@gmail.com} } 

\begin{abstract}
We analyze evolutionary dynamics in a confluent, branching cellular population, such as in a growing duct, vasculature, or in a branching microbial colony.  We focus on the coarse-grained features of the evolution and build  a statistical model that captures the essential features of the dynamics. Using simulations and analytic approaches, we show that the survival probability of strains within the growing population is  sensitive to the branching geometry: Branch bifurcations enhance survival probability due to an overall population growth (i.e., ``inflation''), while branch termination and the small effective population size at the growing branch tips increase the probability of strain extinction.  We show that the evolutionary dynamics may be captured on a wide range of branch geometries parameterized just by the branch diameter $N_0$ and branching rate $b$. We find that the survival probability of neutral cell strains is largest at an ``optimal'' branching rate, which balances the effects of inflation and branch termination. We find that increasing the selective advantage $s$ of the cell strain mitigates the inflationary effect by decreasing the average time at which the mutant cell fate is determined. For sufficiently large selective advantages, the survival probability of the advantageous mutant decreases monotonically with the branching rate.   \end{abstract}


\end{frontmatter}


%
%
\section{\label{sec:intro}Introduction}

Branched cellular populations are found across the tree of life \citep{branchreview}, and especially in organs of animals including the kidney, lung, and mammary glands \citep{IPaine,branchgeo}. Cancerous invasions may also develop branched structures as they invade these tissues  \citep{Breastcancer1,Lungcancer1}.    Root networks and plant vasculature may also branch, creating substrates for subsequent invasions of microbial biofilms \citep{rootbiofilm}.  Certain microbial colonies, such as those formed by \textit{P. aeruginosa},   grow in branched structures under nutrient-limited conditions. The dendritic growth allows for a large actively-growing cellular population compared to a uniform colony edge, thereby optimizing the colony growth \citep{aeruginosa}. Certain  corals are  also  branching cellular populations, with growth confined to the branch tips  \citep{coral1,coral2}.  On a larger scale, animal species are sometimes confined to grow and evolve on branching geometries provided by cave networks and streams, for example  \citep{dendriteeco}. Such branching populations provide a unique environment for evolutionary dynamics. In this paper we study the simplest consequences of  branching growth by calculating the survival probability of cellular strains growing within a branching, confluent tissue.

 We focus in this paper on confluent, branched cellular populations such as those found in animal organs including the lungs and kidney ducts.  In animals, this   branching morphogenesis  is complex and involves cell migration and molecular regulation \citep{aoespinosa}. Here we take a coarse-grained approach and consider the basic structure of such populations and the consequences for evolution. Recently,   \cite{simons1} have demonstrated that a branching and annihilating random walk model captures the major geometrical features of such branching populations. We will build our branching population model based on their results, extending the model to allow for an actively-dividing cellular population with strain-strain competition at branch tips.  In such tip-driven growth, the cells at the branch tips will be the ones that contribute to the evolution of the population. These populations are  examples of ``range expansions'' in which a population grows into a new territory. The geometry and spatial structure of the range expansion has a profound influence on its evolutionary dynamics \citep{KorolevRMP,ExcoffierReview}, which we will also find for the branching structures.

The salient properties influencing survival probability of cellular strains within the branching structure are the dividing population size  increase due to branching, the extinction of dividing populations due to branch termination,  and the selective advantage $s$ of strains within the dividing population.  We will calculate the survival probability of a mutant strain within the branching structure, taking into account all three major properties. We will also pay special attention to the survival of selectively neutral $(s=0)$ strains, showing how the branch rate $b$ and the branch tip population size $N_0$ impact the survival of cell lineages within the population.

The basic structure of the branched populations will be a (self-avoiding) branching and annihilating random walk (BARW). These BARWs  are interesting in their own right, as they describe other physical phenomena including  catalytic reactions \cite{reactionBrowne} and the configurations of branched polymers. For the latter, self-avoiding BARWs create structures which can approximate randomly branched polymers and the excluded volume interactions (i.e., self-avoidance) are particularly important for understanding, for example, the scaling of the polymer radius of gyration  \citep{polymerExcludedVolume,polymerExcludedVolume2}. In our case, we will also focus on self-avoiding BARWs, which capture the basic geometric features of branching cellular populations, as recently verified by comparisons with branching tissue reconstructions \citep{simons1}. The branch ``annihilation'' in this context refers to the termination of growth at branch tips which grow within some specified distance to an extant branch. This termination is a  general feature of self-avoiding, branched cellular populations. For example, such branch ``annihilation'' occurs in diffusion-limited growth of microbial colonies where the branch tips cannot grow near an existing portion of the colony due to nutrient depletion \citep{DLAbacteria}.

The spatial structure of a population can have a profound effect on its evolutionary dynamics. In the tip-driven growth considered here, the dividing cell population is small. This means that genetic drift (i.e., small number fluctuations) is enhanced and mutants occurring within the dividing cell population may be driven to extinction via such fluctuations. In other spatially distributed populations, such as microbial colonies grown on a Petri dish,   local extinction and fixation leads to the formation of ``genetic sectors'' where single strains locally fix \citep{hallatschekPNAS,ExcoffierReview}. Such sectoring may be possible within individual branches, as well, as long as the cells in the branch maintain a confluent structure with little cell rearrangement. A schematic of a genetic sector on a cylindrical branch is shown in Fig.~\ref{fig:intro}(a). Such sectors have been studied extensively with various models including the stepping-stone model \citep{KorolevRMP}, cellular automata approaches \citep{MKNPRE}, and more explicit, individual-based simulations of cell division \citep{WaclawMech}.

On top of the effects of genetic drift, the curvature of the population front will also impact strain survival. For example, when the total population size increases with time, such as in a growing cluster of cells at the expanding  edge of a curved microbial colony, the genetic drift is suppressed and strains within the population are more likely to survive  \citep{MKNPRE,MOLSph}.  Such population ``inflation'' occurs at bifurcation points in our branching model, as shown schematically in Fig.~\ref{fig:intro}(b). Although  branching events are expected to enhance survival probability, the creation of additional branches increases the probability of branch annihilation as branches become more crowded. For crowded branching structures, the dynamics will be complicated, with strains surviving as long as they are fortuitous enough to land on the branches which continue growing [branches with stars in Fig.~\ref{fig:intro}(c)]. We shall see that branch annihilation dominates at large branching rates, complicating the evolutionary dynamics of strains in branching populations. Our study  establishes some general principles for understanding the evolutionary dynamics of cellular populations constrained by branching geometries. This is important not only for understanding cellular populations which branch naturally, but also invading populations constrained  by existing branched structures, such as ductal carcinomas \citep{tumorbranch}. In this case, it is especially important to understand the survival of strains within the population, as cancers often contain many cell strains, whose survival is strongly influenced by both selection and genetic drift \citep{tumorhet,tumorhetreview}.

The paper is organized as follows: In the next section, we discuss our model of branching cellular populations and establish some general principles for understanding evolution in branched structures. In Section~\ref{sec:branching} we characterize the model-generated structures and show the wide range of accessible geometries. Then, in Section~\ref{sec:SBranch} we characterize the survival probability enhancement due to branching. In Section~\ref{sec:SBARW} we include branch termination and discuss survival on fully-developed branching tissues with branch bifurcation and annihilation. We show that for neutral cell strains, there is an optimal branching rate which yields a survival probability maximum. For strains with a sufficiently large selective advantage, the survival probability decreases monotonically with branching rate due to branch collisions. We discuss our results and draw some general conclusions in Section~\ref{sec:Conclusions}.

\section{\label{sec:model} Modelling branching structures}

 \begin{figure}[ht]
        \centering
        \includegraphics[width=0.35\textwidth]{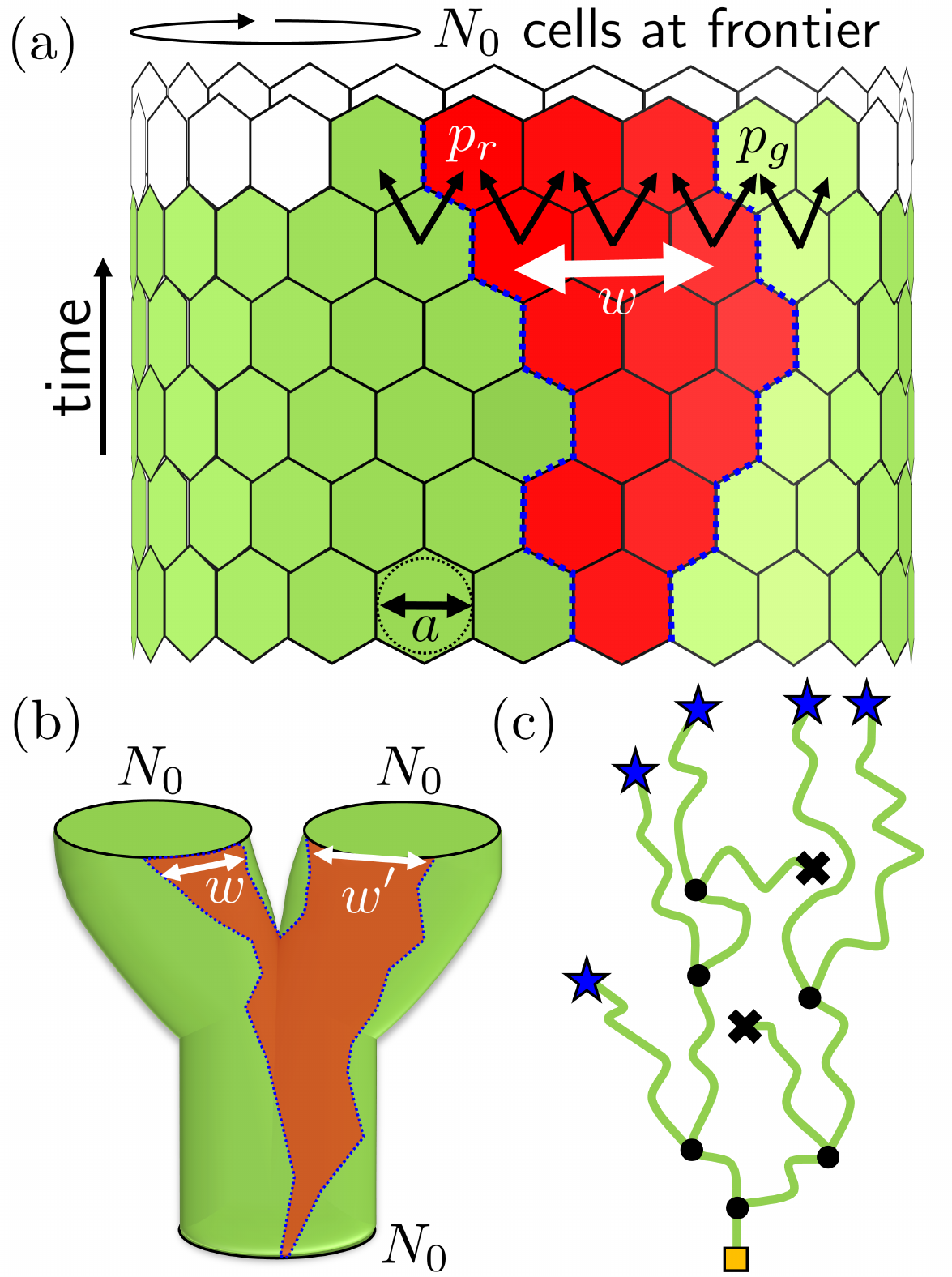}
        \caption{(a) Red and green cells with diameter $a$ (indicated on one green cell) growing as a single tube on a triangular lattice, with a single red cell in a total population of $N_0$ otherwise green cells. The tube grows one generation at a time, and two cells compete to divide into empty spaces at the frontier  as indicated by the short black arrows.  The probability that the red (green) cell wins and divides is $p_r=1/2+s$ ($p_g=1/2-s$).  To calculate survival probabilities, we track the motion of the green/red cell interfaces (blue dashed lines) which bracket a mutant ``sector'' of size $w$.  (b) A schematic of population evolution on a single bifurcating tube. Note that the size of the frontier doubles before a branch, from $N_0$ to $2N_0$. The mutant sector sizes $w$ and $w'$ on the two branches may, in general, be different. (c)  Simplified schematic of our simulated structures with bifurcating (black disks) and annihilating (black crosses) branches. We calculate the survival probability of a single red mutant cell that first appears at the root of the structure (orange square). The branch annihilations will suppress the survival probability. The mutant has to grow on a surviving branch tip (stars) to survive. \label{fig:intro} }
\end{figure}

We begin by describing our model of strain-strain competition at a growing branch tip. We consider a single mutant cell on a linear domain with periodic boundary conditions, i.e., the branch circumference at the tip. During each generation time $\tau_g$, the tip grows by one cell diameter $a$, filling in a new generation of $N_0$ cells along the branch circumference, forming a cylindrical structure as shown in Fig.~\ref{fig:intro}(a).  After multiple generations, the branch cylinder becomes tiled with cells [which we represent with hexagons in Fig.~\ref{fig:intro}(a)]  arranged in a triangular lattice.  As shown in Fig.~\ref{fig:intro}(a), each such step involves pairs of adjacent cells competing to divide into an  empty location at the frontier, such that each cell has a chance to divide into two empty spaces [black arrows in Fig.~\ref{fig:intro}(a)]. If the pair of adjacent cells is the same color, then the empty site is filled with a cell of the same color. Otherwise, if the pair of cells consists of a green ``wild-type'' and a red mutant, then the probability the new cell is green is $p_g=\frac{1}{2}-s$ (or red with probability $p_r=\frac{1}{2}+s$), where $s$ is a selective advantage enjoyed by the red mutant strain. All empty cells at the frontier have to be filled before the branch tip can advance.

As we have pairs of cells competing to divide and non-overlapping generations, our model combines aspects of the Moran \citep{Moran} and Wright-Fisher models of well-mixed population genetics, adapted to this spatial setting. More specifically, it is a small local population size limit of the stepping stone model (see, e.g.,  \cite{KorolevRMP} for a review).  More details about the evolutionary dynamics model are given  in  \cite{MKNPRE}.  These dynamical rules generate a cluster of mutant red cells with boundaries [blue dashed lines in Fig.~\ref{fig:intro}(a)] that perform biased random walks, with the bias driven by the selective advantage $s>0$.  Note that after each generation time $\tau_g$, a boundary will move a distance $a/2$ toward the green side with probability $p_r$ and toward the red with probability $p_g$. Thus, the bias of an individual boundary is   $v_b =(p_r-p_g)a/2 \tau_g= as/\tau_g$. Moreover, if \textit{both} boundaries bracketing the red mutant sector [see Fig.~\ref{fig:intro}(a)] move either toward or away from the red sector, the width $w$ of the mutant sector will change. The resultant bias in the growth of $w$ is thus given by $v=(p_r^2-p_g^2)a/\tau_g= 2as/\tau_g=2v_b$. Along with the bias of boundary motion toward   the green regions when $s>0$, the boundaries will also diffuse (as a consequence of genetic drift). The appropriate diffusion constant $D$ for these interfaces (for small values of $s \ll 1$)  may be calculated from the mean squared displacement of the boundaries during each generation time. Each individual boundary will have a diffusion coefficient given by: $D_b=(p_g+p_r-(p_r-p_g)^2)a^2/8\tau_g\approx a^2/(8\tau_g)$, which yields a diffusion coefficient $D =2D_b\approx a^2/(4 \tau_g)$ for the width $w$ of the mutant sector (see \cite{MKNPRE} for more details). We use this particular model of cell divisions and competition  for computational simplicity, but expect the important features of the survival probability to not depend on the precise choice of microscopic dynamics. Previous work with related models \citep{MKNPRE,MOLmeltdown} shows that the lattice approach reproduces the coarse-grained features of  a range of possible microscopic dynamics. Note that as long as the population locally fixes, other dynamical models may be mapped to our model by an appropriate choice of effective $D$ and $v$.

The long-time survival probability $P_{\mathrm{surv}}$  for a  mutation on a single branch [red region in Fig.~\ref{fig:intro}(a)] follows from the first-passage properties of the blue interfaces shown schematically in Fig.~\ref{fig:intro}(a): If the two blue interfaces meet before completely wrapping the cylinder, then the red mutant strain goes extinct. Otherwise, the red strain will completely take over the branch.   The standard  first-passage result (see, e.g.,  \cite{redner}) reads
\begin{equation}
P_{\mathrm{surv}}= \dfrac{1- e^{-v n_0a/D}}{1 - e^{-v N_0a/D}}  \approx\dfrac{1- e^{-8s  }}{1 - e^{-8N_0s }},  \label{eq:kimura}
\end{equation}
where $n_0=1$ is the initial number of red mutant cells. Apart from a replacement of $D$ with a diffusion constant associated with small number fluctuations (genetic drift), the formula in Eq.~\eqref{eq:kimura} is identical to Kimura's celebrated formula for well-mixed populations \citep{kimura}, where $N_0$ would be the total population size.  In fact, this formula is remarkably insensitive to the particular geographic structure of the population, as noticed some time ago by \cite{maruyama}.

We note that   Eq.~\eqref{eq:kimura} does not apply in certain cases when the position of a cell impacts its ability to propagate, as found in  ``non-isothermal'' structures discussed in  \cite{isothermal}.  Also, when the strain has a large selective advantage $s$ at a branch tip, it may generate a bulge and modify the structure of the branch. The modified structure may, in turn, influence the survival probability. These structure changes will depend on physical interactions between strains and the local environment, as observed in microbial range expansions \citep{nelsonphysical}. In our model, we do not consider the feedback of the evolutionary dynamics on our branching structure or physical interactions between strains. It would be of interest to include such details in the future. Nevertheless, a broad class of cell rearrangements, including mixing due to nearby cell exchange on our cylindrical branches [see Fig.~\ref{fig:intro}(a)], does   not change the survival probability given by Eq.~\eqref{eq:kimura}. A variation of Eq.~\eqref{eq:kimura} also works in higher spatial dimensions \citep{doering,KorolevRMP} and in the presence of flows \citep{pigolotti1}. This means that we may reasonably expect our results to be robust to specific choices of implementation of the cellular growth at the tips, even if this growth is complicated and includes cell rearrangement  \citep{simons2}.

 We also note that the   ``genetic sectoring'' dynamics we consider here are useful simplifications as they allow for us to simulate large branching structures and to make analytic predictions for the survival probability.   Our cell generations are conveniently chosen to be a cell diameter $a$ apart along the branch cylinder, so that the overall density of the cellular population on the cylindrical branch surface is $1/a^2$. It is possible to choose other packing densities, such as a close-packed structure, by choosing a different separation ($\sqrt{3}a/2$ for close-packed). This choice does not impact the properties of the survival probability which we study here.

 Next, we consider the population evolution at a single branch bifurcation point, as shown schematically in Fig.~\ref{fig:intro}(b). We model branch bifurcation by assuming that the population size $N(t)$  grows rapidly over a short period of time to double the population size from $N_0$ to $2N_0$ near the parent branch tip.  Specifically, as soon as a branch is slated to bifurcate, the population at the frontier, $N(t)$, grows according to
\begin{equation}
N(t) =N_0\left(1+\frac{\lambda t^2}{\tau_g^2}\right), \label{eq:growthlaw}
\end{equation}
with the time $t \in [0, t_b]$, where $t_b\equiv \lceil \lambda^{-1/2} \rceil\tau_g$ is the time to branch (given as an integer number of generations) and  $\lceil x \rceil$ is the least integer greater than or equal to $x$.  We fix the parameter $\lambda= 1/200$, yielding $t_b=15 \tau_g$. In this scheme, new branches will always have the same population size as the parent branches. The model easily accommodates other possibilities, including thinning or thickening of daughter branches and different bifurcation region sizes.  Equally-sized branches are common in many structures, including capillaries networked through tissues \citep{LessJ}. When adding cells to the population, a slot between two adjacent cells is chosen at random with the new cell's strain decided by competition between the two (previously adjacent) cells. After the population growth to a total size $2N_0$ [at time $t_b$], a random cut is introduced which splits the population into two equal halves representing the two daughter branches. The two halves then grow independently as cylindrical branches. Note that our choice of a single cut ensures that the two branches each have at most one mutant sector, allowing for some computational simplicity in keeping track of the state of each growing branch tip.  
The details of how the strain-strain competition is modelled during the branch growth are given in \ref{appx:modeldetails}. 

We choose a quadratic growth function $N(t)=N_0(1+\lambda t^2/\tau_g^2)$ because this function generates smooth bifurcation regions. Note that $\frac{dN}{dt}=2 \lambda N_0 t/\tau_g^2$ vanishes at $t=0$ for this function choice, so that the cylindrical branch transitions into the bifurcation region smoothly. This corresponds to a gradual ramping up of the cell growth rate in the branch bifurcation region. Other possibilities, such as $N(t)=N_0(1+\lambda t/\tau_g)^2$, generate a cusp at the junction between the cylinder and bifurcation region. One may consider a range of possibilities for $N(t)$ as it is likely that different branched structures will have different $N(t)$\ that better approximate the cell growth in the branch bifurcation regions. We find, using the analytic techniques described in Section~\ref{sec:SBranch}, that  the choice of $N(t)$ does not significantly modify the survival probability of a mutant strain as long as the time $t_b$ it takes to double the population from $N_0$ to $2N_0$ (to generate two new daughter branches) remains fixed. Then, there is less than 5\% difference in survival probability between our results for the quadratic growth function and for various other choices including   $N(t)=N_0(1+\lambda t/\tau_g)^2$, $N(t)=N_0(1+\lambda t/\tau_g)$, and $N(t)=N_0 e^{\lambda t/\tau_g}$.

To create a complete branching structure, we allow for a  cylindrical parent ``stalk'' to  branch at a fixed rate $0 < b \leq 1$. Specifically, after each generation, there is a probability $b$ that the cylindrical branch begins to bifurcate according to the growth law in Eq.~\eqref{eq:growthlaw}. Thus, we have a branching rate $b/\tau_g$, with $\tau_g$ the generation time (corresponding to an average time $\tau_g/b$ before a cylindrical branch will bifurcate). The parameter $b$ will influence the branch geometry as the structure evolves. For example, higher branch rates $b$  are common in interconnected capillary networks, while smaller $b$ values occur for longer ducts like those of arteriolar networks \citep{LessJ}. For a fixed probability $b$, the average length of the cylindrical portions of the branching structure is $\ell_b\sim a/b$, with $a$ the cell diameter.~As we also fix the population size $N_0$ of the branch tips, our structures have constant values of the dimensionless ratio   $\ell_b/(aN_0)$. For human lung tissue, for example, $\ell_b/(aN_0)\approx 1$ \citep{branchgeo}. We will consider similar values for our structures.

  We choose the daughter branches to split evenly away from the parent, in a plane with a normal perpendicular to the parent branch growth direction. This is called planar or orthogonal bifurcation and is distinct from lateral branching where daughter branches grow off from the side of the parent branch \citep{branchgeo}. The planar bifurcation is common in branching tissue found in kidneys and mammary glands \citep{IPaine} and is driven by tip cells. Lateral branching  is common in blood vessels and the dendrite stalks on nerves. This branching is often driven by single cell extension, typically in response to stimuli \citep{buildingbranched}. We also allow for branch wandering and for branch death due to crowding, where a collision between a growing branch tip  and any existing branch structure will result in the termination of the growth. The midlines of our branches evolve according to the dynamics described by \cite{simons1}.

\begin{figure}[ht]
        \centering
        \includegraphics[width=0.4\textwidth]{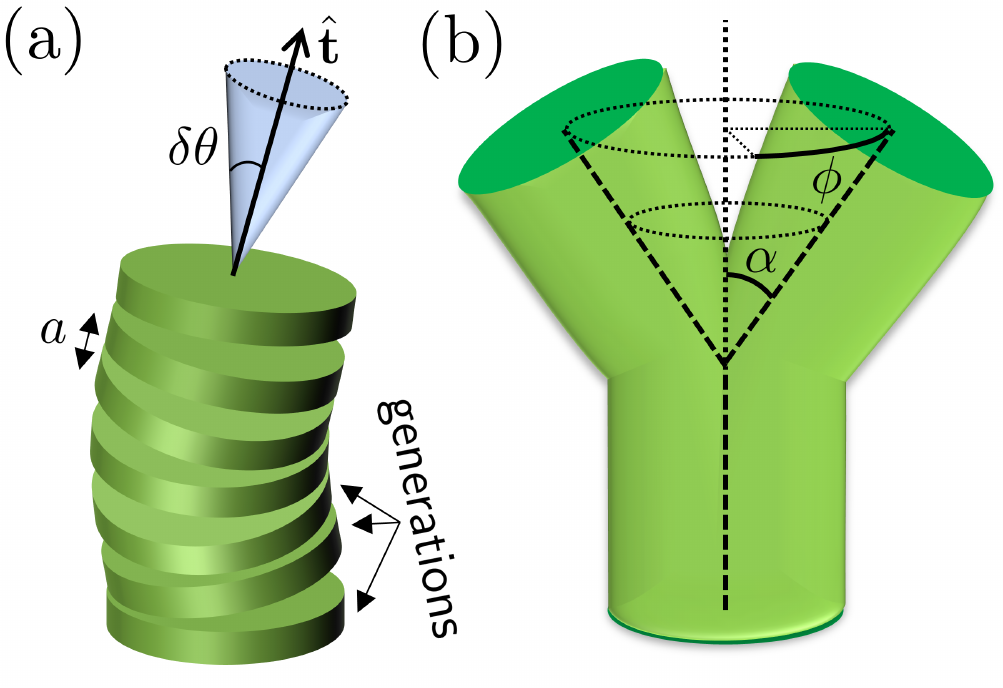}
        \caption{ (a) A schematic of a growing branch. For each generation along the branch (solid disks), the branch undulates by changing the growing direction $\hat{\mathbf{t}}$ by a random direction inside the indicated cone, with angular size $ \delta \theta = 2 \pi/N_0$. Once the new random direction is chosen, the branch advances in that direction by a cell diameter $a$. (b)   A schematic of a bifurcation event in which  the two daughter branches grow in a random direction in a cone centered around the direction of the parent branch. The cone angle $\alpha$  is a uniformly-distributed random number between $\alpha_0 \pm \delta \alpha$, where $\alpha_0=5\pi/18$ and $\delta \alpha / \alpha_0 = 1/3$.  \label{fig:BARWsetup}}
\end{figure}

 To allow for branch undulation as the branches grow through what is likely a rather heterogeneous environment, we allow for the orientation of each branch to change stochastically with each generation, following the branching model proposed by   \cite{simons1}: Every generation time $\tau_g$, the tip of every branch rotates away from the current direction of propagation by a random angle $\delta \theta \in [0,2 \pi/N_0]$ and at a random polar angle $\phi \in [0,2\pi)$. In other words, each generation, every growing branch chooses a slightly new direction inside a solid angle of size $2\pi[1- \cos  (\delta \theta)]$. The random directions are conveniently chosen using an efficient algorithm described in \cite{Arvo92fastrandom} for sampling uniform random points on a sphere. This is shown in Fig.~\ref{fig:BARWsetup}(a). Upon bifurcation,  the two daughter branches split off at equal angles away from the   parent branch direction, at azimuthal angles $\alpha_0 \pm \delta \alpha$, where   $\alpha_0 =0.873$ and $\delta\alpha/\alpha_0 = 1/3$. The particular choices for $\delta \alpha$ and $\alpha_0$ only weakly influence the overall branching structure \citep{simons1}. The two daughter branches are oriented randomly in the polar direction, with one branch at $\phi$ and the other at $\pi-\phi$, where $\phi \in [0,2 \pi)$, such that the angle between the two is always $2\alpha$. The geometry is illustrated in Fig.~\ref{fig:BARWsetup}(b). Individual branches, then, will behave as elastic ``chains''  or ``polymers'' with persistence lengths depending  on the choice of $N_0$, which fixes the range of the random angle $\delta \theta$. 

We terminate any growing branch tip if it comes within a distance $d_a = aN_0/2\pi$ of any other other branches (including the tip's own branch stalk). More details of this procedure, including special considerations at the branch bifurcation regions, are given in \ref{appx:modeldetails}. Our structure is thus an expanding system of self-avoiding, branching, annihilating random walks representing the branch tips. The two parameters necessary to generate the structure are  $b$ and $N_0$, with the latter determining both the annihilation distance  $d_a = aN_0/2\pi$ and the angular  magnitude of branch undulations $\delta \theta=2\pi/N_0$. Just these two parameters allow for a broad range of structures, as shown in the phase diagram in Fig.~\ref{fig:branchpics}.  We may find structures ranging from dense, tortuous branches at small $N_0$ and large $b$, to thick, loosely packed straight branches at large $N_0$ and small $b$. Some examples of the structure variety are shown in Fig.~\ref{fig:branchpics}(a)-(d). Note that in some branching cell tissues, the branching is driven by single cells \citep{buildingbranched}. Our model does not capture this interesting method of bifurcation as the $N_0\sim 1$ limit does not allow for cell competition. Instead, our model focuses on  branching involving a larger group of actively-dividing cells at the branch tips. Lung epithelia, for example, would be consistent with groups of tens or more cells at the tips \citep{lungbranch}. Here we will focus on $N_0\sim 50-100$, as we would expect the cell-level details of the branch bifurcation to be less important for these somewhat larger cell populations. The branching model code and the raw data from the simulations are readily available online \citep{github}.  
  
\section{\label{sec:branching} Branching structure characterization}

\begin{figure}[ht]
        \centering
        \includegraphics[width=0.45\textwidth]{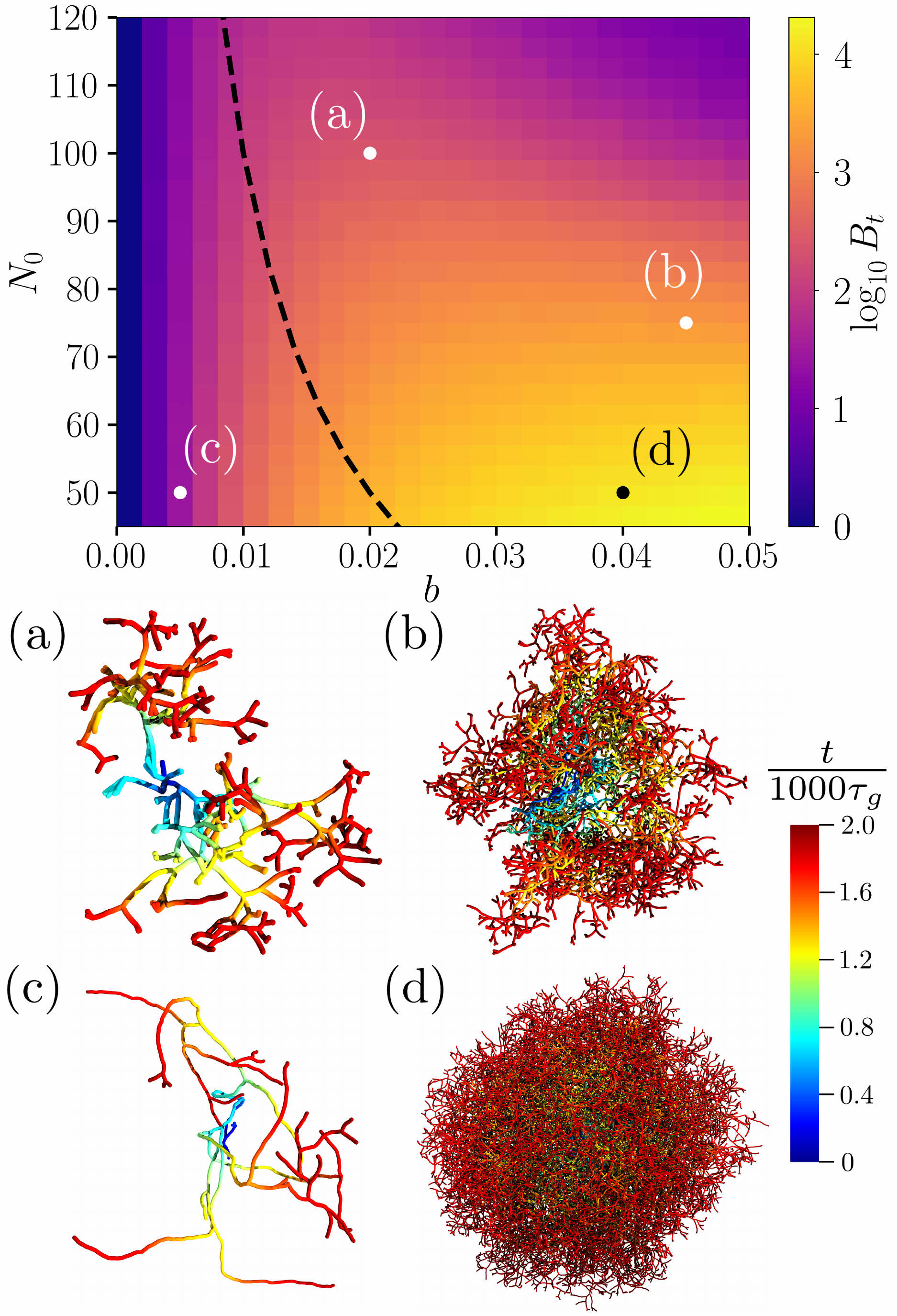}
        \caption{ Number of terminal branches $B_t$ after 1000 generations as a function of the branching rate $b$ (in inverse generation times $\tau_g^{-1}$) and the branch circumference $N_0$ (given in numbers of cells). Examples of the simulated structures are given in (a) through (d) where the colors indicate the time-evolution of the structure (from blue to red). The branch thicknesses are proportional to  $aN_0$, but are drawn thinner than the actual thickness in order to more easily visualize the structure. Note that  at small $N_0$ and large $b$ we find dense packings of branches as the effective branch radius is small. The small $N_0$ also yields a shorter branch persistence length $\xi_p$ which yields more undulated branches as shown in (c). The dashed line shows the boundary $b N_0  =1$. For values of $b$ to the right of this line, branch terminations are prevalent, as seen in structures (a), (b), and (d). 
 \label{fig:branchpics}}
\end{figure}

The value of $N_0$ determines both the annihilation radius  $d_c$ of the branches and the persistence length $\xi_p$.  It is clear that when the branching rate $b$ (per generation time $\tau_g$) becomes large enough that the average branch length $\ell_b$  becomes comparable or larger than $d_a$, then branch terminations will start to dominate the structure. This happens when $\ell_b \sim a/b   \lesssim d_a \sim aN_0 $, or $b \gtrsim  1/N_0$. The line $bN_0=1$ is shown in Fig.~\ref{fig:branchpics} and we see that the structures (a), (b), and (d) all fall into the regime $b > 1/N_0$  where branch terminations will be prevalent. Note that, in our simplified model, the annihilation distance $d_a=aN_0/2\pi$ is proportional to the number of dividing cells $N_0$, a restriction which may be lifted in a more detailed model. For example, it is likely that a diffusive cell signal may arrest the growing branch tip before it comes within some specific distance of an existing branch. This might necessitate an independent value $d_a > a N_0/2\pi$. Such a generalization may introduce different kinds of branching structures, and may be easily incorporated in our model. Here we focus on a fixed $d_a=aN_0/2\pi$ for simplicity and consider just two branch parameters $b$ and $N_0$, which already capture a wide range of branch geometries analogous to those found in natural cellular populations.

\begin{figure}[ht]
        \centering
        \includegraphics[width=0.45\textwidth]{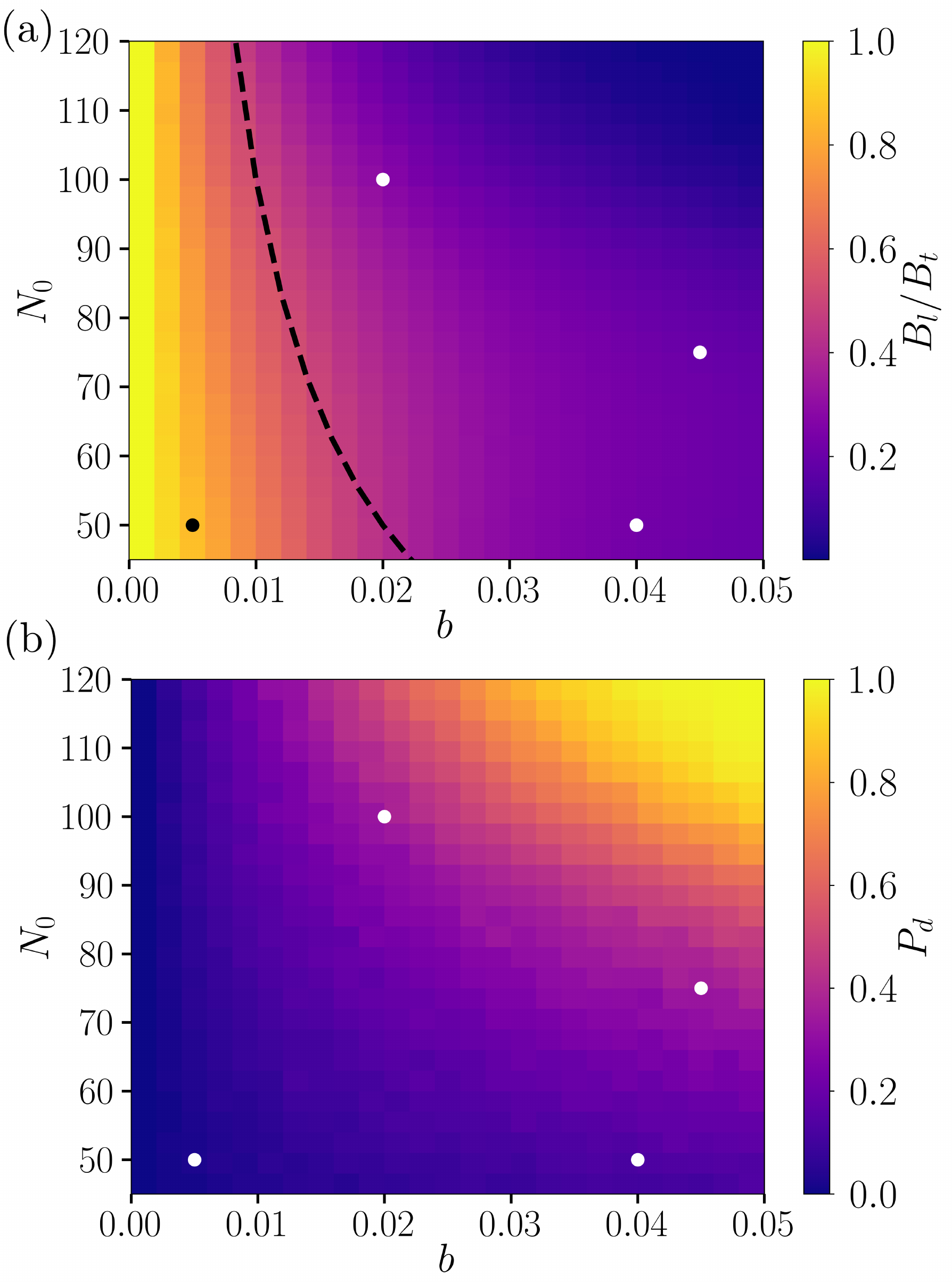}
        \caption{ (a) Ratio $B_l/B_t$ of the ``living'' terminal branches $B_l$ (branch tips which have not collided) to the total number of branch tips $B_t$ after 1000 generations of branch structure evolution. The dots indicate the same positions shown in Fig.~\ref{fig:branchpics} where examples of surviving branched structures are shown. We see that at large $N_0$ and $b$ (given in units of inverse generation times $\tau_g^{-1}$), the fraction is significantly smaller as the branching structures have many terminated side branches, as shown in Fig.~\ref{fig:branchpics}(a,b). The dashed line indicates the condition $N_0 b=1$, separating the two regimes in which branch termination is more or less prevalent.  In (b) we calculate the probability $P_d$ of the entire structure self-annihilating with each branch tip colliding after 1000 generations. Note that a significant probability of structure ``death''  occurs at large $N_0$ and $b$. The white dots correspond to the same examples shown in Fig.~\ref{fig:branchpics}.      \label{fig:branchphases}}
\end{figure}

For smaller $N_0$, we expect more undulated branches that rarely run into each other. Conversely, at larger $N_0$, we find more straight, thick branches and frequent branch collisions. So, at large $N_0$ we find structures with many terminating side branches, as shown in Fig.~\ref{fig:branchpics}(b). The effects of these annihilations can be characterized by studying the fraction $B_l/B_t$ of free, growing ends $B_l$ to the total number of branch ends $B_t$. We see in Fig.~\ref{fig:branchphases}(a) that this ratio decreases rapidly with increasing $N_0$   and $b$. Indeed, at sufficiently large $N_0$ and large $b$, all of the branches will run into each other and terminate. The condition $N_0b=1$  provides a good estimate of when there will be more branch terminations compared to surviving branches, as can be seen in Fig.~\ref{fig:branchphases}(a). The probability $P_d$ of the entire structure self-annihilating for a given instance of the structure evolution is shown in Fig.~\ref{fig:branchphases}(b). We get a significant probability only when both $b$ and $N_0$ are large. In this regime, we find stubby tree-like structures.

Since we are interested in survival probability on extended branched structures, we will avoid the problem of complete structure death by only looking at ensembles of branching structures which have survived up to a fixed generation time (1000 generations for our simulations). This ensures that the survival probability of mutations in the structure only depend on the behavior of the side branches and not the survival probability  $P_d$ of the entire branching structure, shown in Fig.~\ref{fig:branchphases}(b).  This choice of ensemble is necessary as completely avoiding the total branch structure death is not possible for all values of $N_0$ and $b$ we are interested in. The total branch structure death is an undesirable condition for our analysis as mutations may fix rapidly and completely take over the branching structure, especially at large selective advantages $s$. We want to ensure that such cases are properly treated as survival events for the mutations. The total branch structure death also complicates our analytic approaches to this problem, as discussed in more detail in the next sections.

\begin{figure}[ht]
        \centering
        \includegraphics[width=0.45\textwidth]{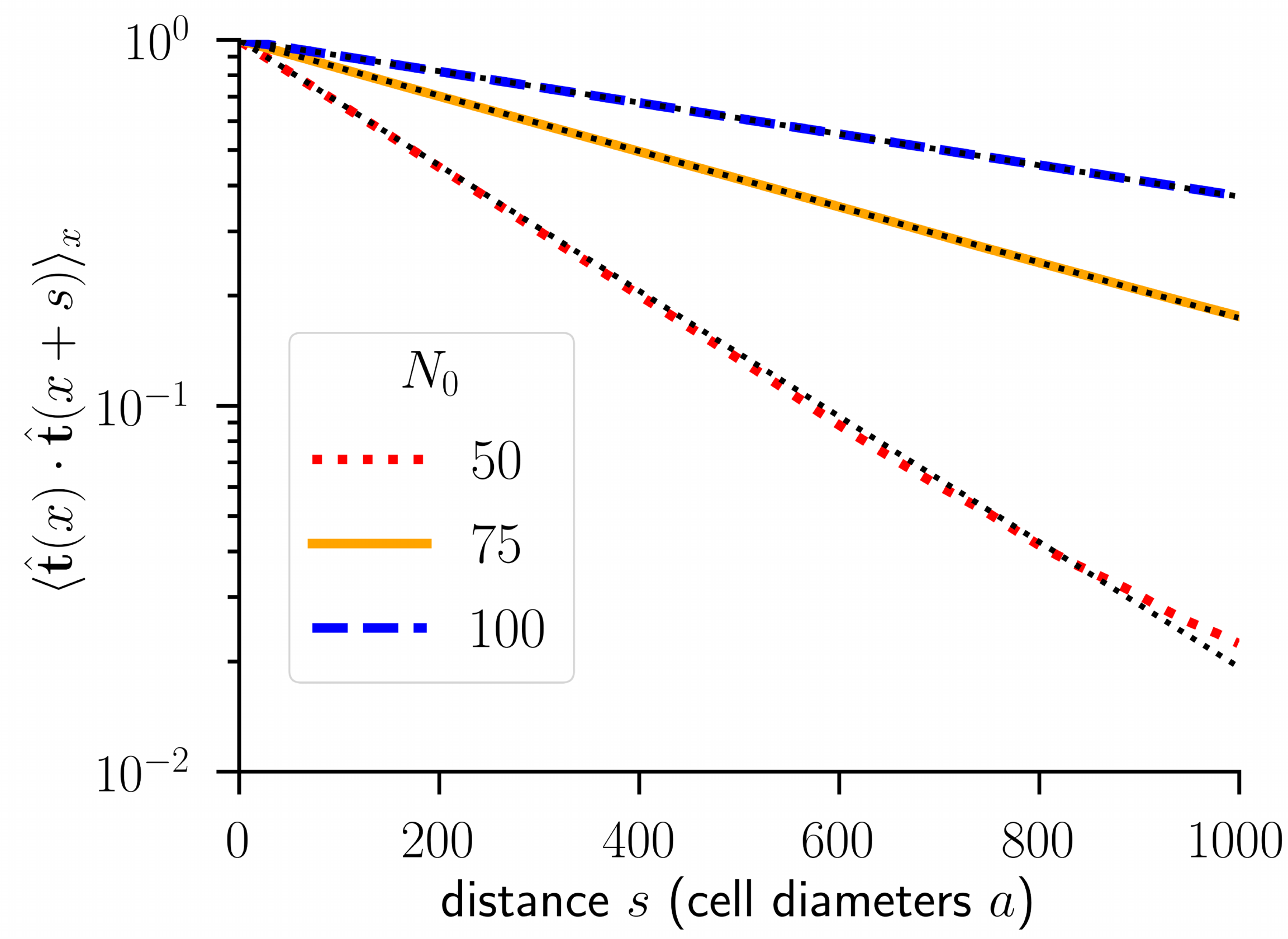}
        \caption{ Orientation correlation decay as a function of distance $s$ along a single, undulating branch, calculated from averaging over 100 single-branch simulations with branch lengths of 3000 generations. Note that the circumference of the branch strongly influences the decay of the correlation (i.e., the persistence length).
The black dotted lines are fits to the persistence length $\xi_p$ defined via: $\langle \hat{\mathbf{t}}(x) \hat{\mathbf{t}}(x+s)\rangle_x = e^{-s/\xi_p}$. We find $\xi_p \approx250 $ for $N_0=50$, $\xi_p \approx570 $ for $N_0=75$, and $\xi_p \approx1000 $ for $N_0=100$. Note the clear exponential decays found in the simulations, as may be expected from similar models describing orientation correlations in polymers  \citep{kratky}.
 \label{fig:persistence}}
\end{figure}

We now  briefly mention the structure of the \textit{individual} branches. Our update rules  generate branch undulations which may be characterized by a persistence length $\xi_p$, which we now describe. We may analyze the branch structure by turning off branch annihilation and branching $(b=0)$ and measuring the orientational correlations $\langle \hat{\mathbf{t}}(x) \hat{\mathbf{t}}(s+x)\rangle_x$ of a single branch [see Fig.~\ref{fig:BARWsetup}(a)], averaging along the branch length $x$. The results are shown in Fig.~\ref{fig:persistence} for different values of $N_0$, the single parameter determining the  branch structure. We   find a clear exponential decay for the correlation (see black dashed lines in Fig.~\ref{fig:persistence}), with a characteristic scale $\xi_p$ which defines the persistence length. This is expected because turning off the branch collisions reduces our model to a polymer model \citep{kratky},  with the generations in our model representing discrete monomeric units attached in a ``chain''. The resultant  structure takes the form of a persistent random walk (see, e.g., \cite{kratkyrandom}), in which discrete segments tend to be aligned in the same direction but eventually achieve a random walk on length scales larger than $\xi_p$. Note that varying $N_0$ has a pronounced effect on the  correlation decay, with smaller $N_0$ yielding a much smaller persistence length $\xi_p$. Including collisions would modify the orientational correlations as excluded volume effects are known to  modify the exponential decay \citep{kratkyexcludedvolume}. Nevertheless, the persistence length $\xi_p$ calculated here provides a valuable characterization of the branched structure.  For example, we see in Fig.~\ref{fig:persistence} that structures with $N_0=100$ will have very straight branches, with very little undulation even after a thousand generations. Conversely, branches with  $N_0<50$ reorient already after a couple hundred or fewer generations. 

\section{\label{sec:SBranch} Survival on branching structures}

We first consider mutant survival on branching structures without any branch collisions. Suppose $N_T(t)$\ is the \textit{total} actively dividing population at time $t$, which starts with an initial ring of $N_T(t=0)=N_0$ cells.  The population then grows  into a cylindrical shape which eventually bifurcates, as shown in  Fig.~\ref{fig:intro}.  For a mutation competing within a total population of size $N_T(t)$, the  probability $P(w,t)$ of observing the mutant forming a genetic sector of size $w$ at time $t$ (see Fig.~\ref{fig:intro}) is given by 
\begin{equation}
\partial_t P(w,t) = \frac{D N_0^2}{[N_T(t)]^2} \partial_w^2 P(w,t)- \frac{v N_0}{N_T(t)} \partial_w P(w,t), \label{eq:diffusion}
\end{equation}
where $D$ is the genetic drift strength and $v$ the deterministic bias, due to selective advantage, acting on the mutant sector of size $w$ \citep{MOLSph,MKNPRE}. In our lattice model (see Section~\ref{sec:model}), $D \approx a^2/(4\tau_g)$ and $v \approx 2sa/\tau_g$, with $a$ the cell diameter and $\tau_g$ the generation time.  Let us review briefly what happens in the case of a fixed population size $N_T(t)=N_0$, partially covered in  Section \ref{sec:model}. In this case, Eq.~\eqref{eq:diffusion} may be solved  for the survival probability $P_{\mathrm{surv}}$ of a mutant sector using first-passage techniques \citep{redner}. The result for  a single initial mutant cell is given by Eq.~\eqref{eq:kimura}.
 If we additionally assume a sufficiently large actively dividing population $N_0$ such that   $N_0s  \gg 1$, then we find\begin{equation}
P_{\mathrm{surv}} =\dfrac{1- e^{-v  a/D}}{1 - e^{-v N_0a/D}}  \approx1-e^{-8s }. \label{eq:noinflation}
\end{equation}
 Note   that for neutral mutations, as $s \rightarrow 0$, we find $P_{\mathrm{surv}} \rightarrow 1/N_0$ from the first equality in Eq.~\eqref{eq:noinflation}. This result is   sensible as a neutral mutation would have an equal chance of fixing any cell in the population of  $N_0$  actively dividing cells at the branch tip.

 Let us now consider the branch bifurcations. To make progress, we will make a simplification and ignore the geometry of the branch bifurcation, focussing entirely on the increase of the actively growing population size $N_T(t)$. In other words, we will ignore  any splitting of sectors that occurs, such as the $w$ and $w'$ shown in Fig.~\ref{fig:intro}(b),  and assume we can treat the mutant population as a single sector [replacing $w$ and $w'$ in Fig.~\ref{fig:intro}(b) with a single sector of size $w+w'$, for example]. This simplification is motivated by the observation, discussed   in Section~\ref{sec:model},  that the survival probability of a mutation generally does not depend on the details of the geographic structure of the population.

Consider, then, a single ``genetic sector'' evolving in a population of size $N_T(t)$. The bifurcation events will  increase $N_T(t)$ by $N_0$. In our model, this occurs stochastically as the branching occurs at rate $b$ along each growing branch. We will make an additional assumption that we may treat these bifurcations on average. Then, on average, $N_T(t)$ will experience a series of inflating and non-inflating regimes:
\begin{equation}
\frac{N_T(t)}{N_0} = \begin{cases}
 1 & 0<\frac{t}{\tau_g}<\frac{1}{b} \\
2^{n-1}\left[1+\frac{\lambda}{\tau_g^2}\left(t-\frac{n}{b}\,\tau_g\right)^2\right] & \frac{n}{b}<\frac{t}{\tau_g}<\frac{n}{b}+\frac{1}{\sqrt{\lambda}} \\
2^n   & \frac{n}{b}+\frac{1}{\sqrt{\lambda}}<\frac{t}{\tau_g}< \frac{n+1}{b}
\end{cases}, \label{eq:popgrow}
\end{equation}
where, after the initial parent branch growth period $0<t<\tau_g/b$, the integers $n=1,2,\ldots$ label the periods of inflation followed by growth at constant population size. We see here that, on average, after   $n/b$ generations, the total population size $N_T$ must increase from $N_0$ to $2^n N_0$ due to the branching (if we ignore the branch annihilation events). These simplifications allow for an (approximate) analytic solution to the survival probability, which we now discuss. 
 
 \begin{figure}[ht]
        \centering
        \includegraphics[width=0.5\textwidth]{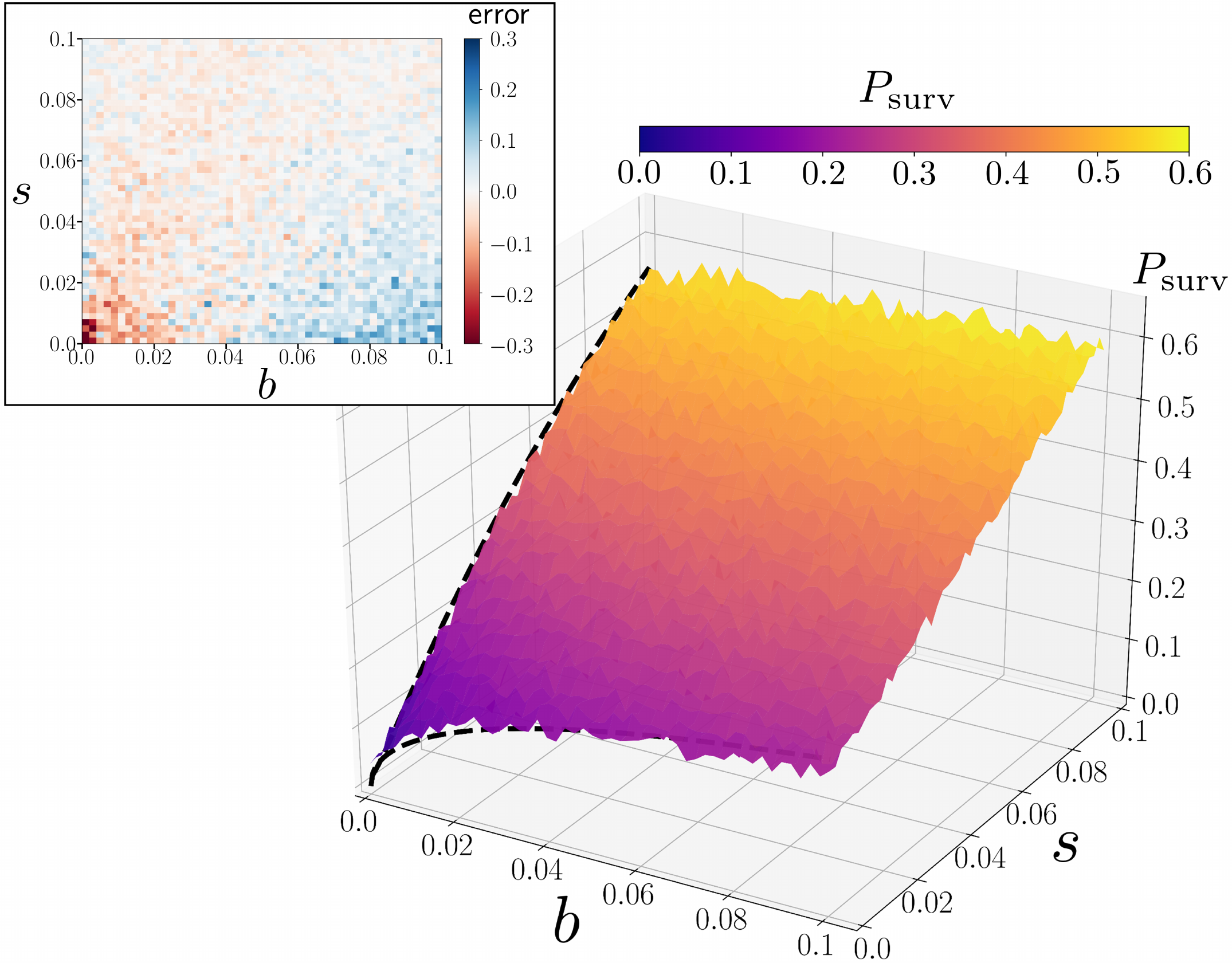}
        \caption{ Survival probability $P_{\mathrm{surv}}$ of a single mutant cell after 1000 generations on a branching structure (with branch collisions turned off) as a function of the mutant selective advantage $s$ and the branching rate $b$. The   surface is generated from simulation results for branch parameter $N_0=75$ .       The black dashed lines show the analytic results for the long-time survival probability in the two important limits $s=0$ [Eq.~\eqref{eq:psurvneutral}] and $b=0$ [Eq.~\eqref{eq:noinflation}]. The inset shows the error $(P_{\mathrm{surv}}^{\mathrm{theory}}-P_{\mathrm{surv}}^{\mathrm{sim.}})/P_{\mathrm{surv}}^{\mathrm{sim.}}$ of our analytic expression  $P_{\mathrm{surv}}^{\mathrm{theory}}$, given by Eq.~\eqref{eq:intsol}, as compared to the simulation results. We see in the inset that although the analytic result simplifies the branch bifurcation dynamics, the comparison between theory and simulations is favorable, with less than 10\% error in most of the phase space.  There are   larger errors in the small $s$ and $b$ regions since the survival probability is small there. We generally expect $P_{\mathrm{surv}}^{\mathrm{theory}}<P_{\mathrm{surv}}^{\mathrm{sim.}}$ in this region as it will take longer than 1000 generations for the probability to approach the theoretical value at small $s$ and $b$.  \label{fig:noannihilations}  }
\end{figure}

The diffusion equation  Eq.~\eqref{eq:diffusion} may be solved   \citep{MOLSph,roughconformal} for an arbitrary time-dependent total population size $N_T(t)$   by introducing a  time-like, dimensionless variable
\begin{equation}
z(t) \equiv \frac{D}{a^2} \int_0^t \mathrm{d}\bar{t} \, \frac{N_0^2}{[N_T(\bar{t})]^2}, \label{eq:z}
\end{equation}
with $a$ the cell diameter and $D$ the diffusion coefficient of the mutant sector width. Transforming from the time $t$  to this new coordinate $z$  removes the time-dependence in the second derivative term in Eq.~\eqref{eq:diffusion}. The velocity term proportional to $v$ remains $z$-dependent, but one may employ an adiabatic approximation, as discussed in more detail in  \cite{MOLSph}, which yields an approximate  solution for the long-time survival probability for a single mutant cell (corresponding to an  initial width $w=a$) at time $t=0$:
\begin{equation}
P^{\mathrm{theory}}_{\mathrm{surv}} \approx 1 - \int_0^{z_m} \mathrm{d}z\, \frac{ \exp\left[ - \frac{1}{4z} \left( 1+ \frac{av N_T[t(z)]}{DN_0}\,z \right)^2 \right]}{2 \sqrt{\pi} z^{3/2}}  , \label{eq:intsol}
\end{equation}
where $z_m = z(t \rightarrow \infty)$ is the maximum value of $z$ as given by the $t \rightarrow \infty$ limit of  Eq.~\eqref{eq:z}.  Note that to perform the integral in Eq.~\eqref{eq:intsol} when $v \neq 0$, one would have to find the time $t$ as a function of the  variable $z$.
We  compare this theoretical result to simulations of our branching structures in Fig.~\ref{fig:noannihilations}, which include all of the details of the branch bifurcation and population splitting. The inset  shows the error between the simulation result and the analytic expression in Eq.~\eqref{eq:z}. Note that the error is less than 30\% over the entire range of parameter space, and typically much smaller. This is remarkable as there are \textit{no fit parameters} used in the comparison.  This means that the details of the population splitting are not as important in determining the strain survival as the population growth that occurs during the branch bifurcation events.  Thus, for actual cellular populations which branch but have very few branch annihilation events, we would expect Eq.~\eqref{eq:z} to describe the survival probability of strains (mutants with a selective advantage $s$, say) independently of the precise details regarding how the branches split. This result is consistent with the known insensitivity of the survival probability of a strain to its spatial distribution within a population that we discussed in Section~\ref{sec:model}. Rather than this spatial distribution, the key determinant of survival is the average  population size $N_T(t)$ at all of the (actively dividing) branch tips.

Let us now consider the case of neutral mutations with $s=v=0$. When $v=0$, the integral in Eq.~\eqref{eq:intsol} may be evaluated exactly and  the full time-dependent survival probability (in the large $N_0$ limit) is given by
\begin{equation}
P_{\mathrm{surv}}(t)=\operatorname{erf}\left[ \frac{a}{2\sqrt{z(t)}} \right], \label{eq:psurvneutral}
\end{equation}
with $z(t)$ defined in Eq.~\eqref{eq:z}.  We may get some important information from this equation by considering a single cylindrical branch, with no bifurcations $(b=0)$ and a fixed $N_T(t)=N_0$. In this case, $z(t)=Dt$. Then, the survival probability $P_{\mathrm{surv}}(t) = \operatorname{erf}[a(4Dt)^{-1/2}]\approx \operatorname{erf}[\sqrt{\tau_g/t}]$ decays as $2\sqrt{\tau_g/\pi  t}$ at long times. So, if we have a cylindrical branch with $N_0$ cells on it, then, on average, the number of strains remaining when the branch bifurcates is $N_0\operatorname{erf}[\sqrt{b }] \approx 2N_0 \sqrt{  b/\pi}$. Even for a small branching rate like $b = 0.001 $ (per generation), we find that over 3\% of the the $N_0$ strains will be present in the population of a branch at the time it bifurcates. We therefore expect inflationary effects to play an important role for these neutral mutations.

The long time survival probability for a neutral mutation in a branching population may be found by taking the limit $z(t \rightarrow \infty)=z_m$ in Eq.~\eqref{eq:psurvneutral}. For the particular growth dynamics in  using Eq.~\eqref{eq:popgrow}, we  find
\begin{equation}
z_m = \frac{D \tau_g}{6b} \left[ \frac{\pi b}{\sqrt{\lambda}}+8 \right].
\end{equation}
Note that when $b \rightarrow 0$, $z_m \rightarrow \infty$, also.
This means that, without branching, apart from a small probability $1/N_0$ that the mutant fixes throughout the entire population, neutral mutations always die out, as expected from the discussion above. When $b>0$, $z_m$ is finite and population inflation at the branch bifurcations can rescue neutral mutations from extinction, similarly to neutral mutation rescue at the surface of an inflating spherical cluster of cells \citep{MOLSph}. Indeed, any non-zero branching rate may rescue a neutral mutation from otherwise inevitable extinction due to genetic drift at large $N_0$. The result for $P_{\mathrm{surv}}$  is given by a black dashed line in Fig.~\ref{fig:noannihilations}. In other words, the ``inflation'' introduced by the population growth at branch bifurcations can force the ``genetic sector'' formed by a mutant strain to grow large enough that the genetic drift is unable to extinguish the strain.

Note that the inflationary enhancement becomes less significant at larger selection coefficients $s$ as the fate of the mutation is determined early in this case. Indeed, either the mutation escapes genetic drift and sweeps the population with its selective advantage $s$, or the genetic drift extinguishes the mutant early on in the evolution (before the branch bifurcation point). We can estimate, using first-passage techniques \citep{redner}, the  characteristic time $t_e$  for extinction of a selectively-advantageous strain: $t_e \sim a/v \sim \tau_g/ s  $, where $\tau_g$ is the generation time and $v$ is the drift velocity induced by the selection [see Eq.~\eqref{eq:diffusion}]. So, we would expect that when $b<s$, the population inflation will not significantly modify the survival probability. This is consistent with our result in Fig.~\ref{fig:noannihilations} as we may see that $P_{\mathrm{surv}}$ does not vary much with $b$ when we consider a fixed, large value of $s$. The fate of neutral mutations ($s=0$), however, will take a longer time to be determined as the extinction or fixation is driven by the slower, diffusive process of genetic drift. In this case, the characteristic time for extinction is $t_e \sim a^2N_0/D \sim N_0 \tau_g $, with $D$\ the diffusion coefficient  [see Eq.~\eqref{eq:diffusion}].  Therefore, for large $N_0$, inflation will play a significant role for the neutral case, as can be seen in Fig.~\ref{fig:noannihilations} where $P_{\mathrm{surv}}$ sharply increases with $b$ at fixed $s=0$.
We will formulate these arguments more precisely in the next section where we consider branch collisions.

\section{\label{sec:SBARW} Survival on branching, annihilating structures}

 We now turn our attention to branch annihilation. These terminations happen when growing branch tips collide with existing portions of the branched structure. To simplify the notation in this section, we adopt time units such that the generation time $\tau_g$ is fixed to unity: $\tau_g=1$. We have  shown in Fig.~\ref{fig:noannihilations} that the details of the population bifurcations are not as important as the increase in the actively dividing population $N(t)$  at each bifurcation event. However, we expect this inflationary effect to compete with branch annihilations which will become more frequent with increased branching rates $b$ [see Fig.~\ref{fig:branchphases}(a)]. It is not obvious in this case that increasing the branch rate $b$ will lead to a larger survival probability. As we observe in Fig.~\ref{fig:branchphases}, increasing the branching rate $b$ at fixed $N_0$ will significantly decrease the ratio $B_l/B_t$  of the number of ``living'' branches $B_l$ on which the mutant may survive to the total number of   branch tips $B_t$. Recall also that in our model there is a chance that the entire branching structure dies out [see Fig.~\ref{fig:branchphases}(b)]. To avoid complications associated with this complete branching structure death, we will here focus on the ensemble of branching structures which survive up to 1000 generations for the purposes of calculating strain survival probabilities.

 \begin{figure}[ht]
        \centering
        \includegraphics[width=0.48\textwidth]{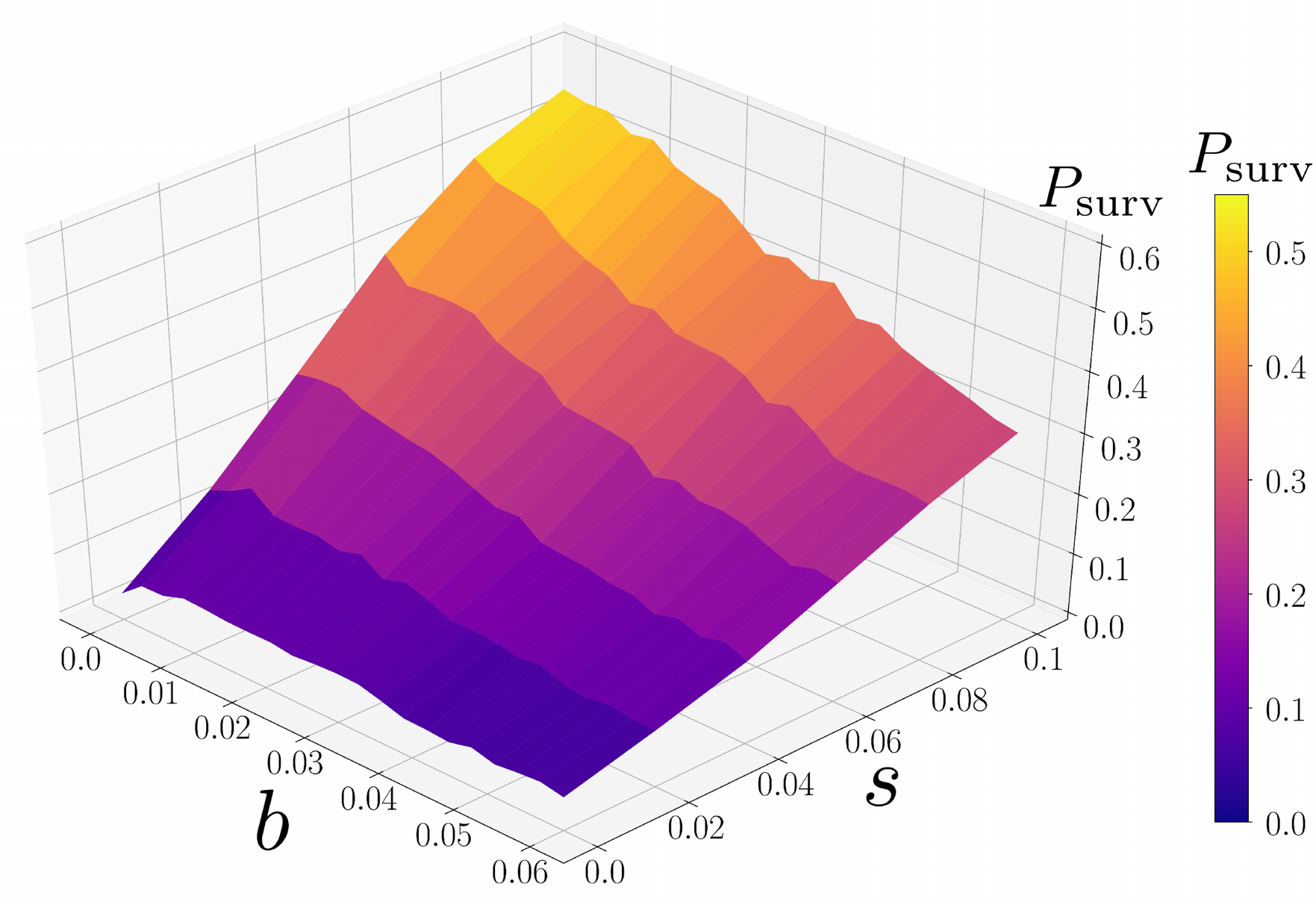}
        \caption{   \label{fig:annihilations} Survival probability $P_{\mathrm{surv}}$ of a single mutant cell after 1000 generations on a branching structure (with branch annihilations) as a function of the mutant selective advantage $s$ and the branching rate $b$. The   surface is generated from simulation results for branch parameter $N_0=75$. Note that, unlike the survival probability on non-annihilating, branching structures (see Fig.~\ref{fig:noannihilations}), the survival probability \textit{decreases} with increasing  branch rate $b$ for larger values of $s$. The survival probability for the neutral case $s=0$ is shown in Fig.~\ref{fig:BARW_bs}.      }
\end{figure}

The results for survival probabilities on the full branching and annihilating structures for $N_0 = 75$ are shown in Fig.~\ref{fig:annihilations}.  The structures are evolved for 1000 generations and we average over an ensemble of 2000 such structures. Comparing these results with Fig.~\ref{fig:noannihilations}, we see that, generally, increasing the branch rate $b$ \textit{decreases} the survival probability when we include branch annihilations. In other words, the deleterious effect of branch annihilations overwhelms the survival advantage a mutation receives from the ``inflationary'' effect due to   the   increase in the dividing cell population at  branch bifurcations. Instead, we find that the survival probability is suppressed because branch annihilation is the dominant effect for most values of $s \gtrsim 0.01$. The exception here is for small selective advantages $s \lesssim 0.01$ where genetic drift dominates the evolutionary dynamics. Here, the inflationary effect from an increasing population size $N(t)$   enhances the survival probability. Let us make a heuristic argument about why we see these results.

One may understand these results by again considering the average time $t_e$ it takes for the mutant on a growing cylindrical branch to go extinct (given that it does do so).   Mutant sectors which survive beyond the time $t_e$  are likely to sweep the population and fix at the branch tips. Using first-passage techniques described in \cite{redner}, we calculate the extinction time
\begin{equation}
t_{e} =\frac{a}{v(\bar{\epsilon}-\epsilon)}\left[\bar{\epsilon}+\epsilon - 2\bar{\epsilon}N_0\left( \frac{ \epsilon-1}{ \bar{\epsilon}-1} \right)\right]   , \label{eq:exttime}
\end{equation}
where $\epsilon = e^{va/D} \approx e^{8s}$ and $\bar{\epsilon}=e^{vaN_0/D}\approx e^{8sN_0}$.
We may consider the behavior of\ $t_e$ in Eq.~\eqref{eq:exttime} at large and small selection strengths:
\begin{equation}
t_e \approx \begin{cases}
\dfrac{4N_0}{3} & sN_0 \ll 1\\[5pt]
\dfrac{1}{2s} &  sN_0 \gg 1
\end{cases}.
\end{equation}
Inflation will only help significantly if the branching rate $b$ is larger than or comparable to $1/t_e$: If the branching rate is much smaller than $1/t_e$, then the fate of the mutant will already be determined (on average) before the branch bifurcation occurs. However, the branching rate $b$ must also be not too large compared to $1/N_0$, because we know from  Fig.~\ref{fig:branchphases}(a)  that the majority of the branch tips will collide and terminate for branching rates much larger than $1/N_0$. So, consider that for $s N_0 \gg 1$, we would require $b \gtrsim 1/t_e \approx 2s \gg 2/N_0$ for the inflationary effect to have much influence on the mutant survival probability.
Note that $b \gg 2/N_0$  is in the regime where we have an overwhelming number of branch terminations. Thus, we conclude that when we have a significant selective advantage, $s N_0 \gg 1$,  branching will have only a \textit{deleterious} effect on the mutant survival probability. This is consistent with the results in Fig.~\ref{fig:annihilations}, where we see that increasing $b$\ will \textit{decrease} the survival probability monotonically whenever $s \gtrsim 1/N_0 \approx 0.01$.

 \begin{figure}[ht]
        \centering
        \includegraphics[width=0.48\textwidth]{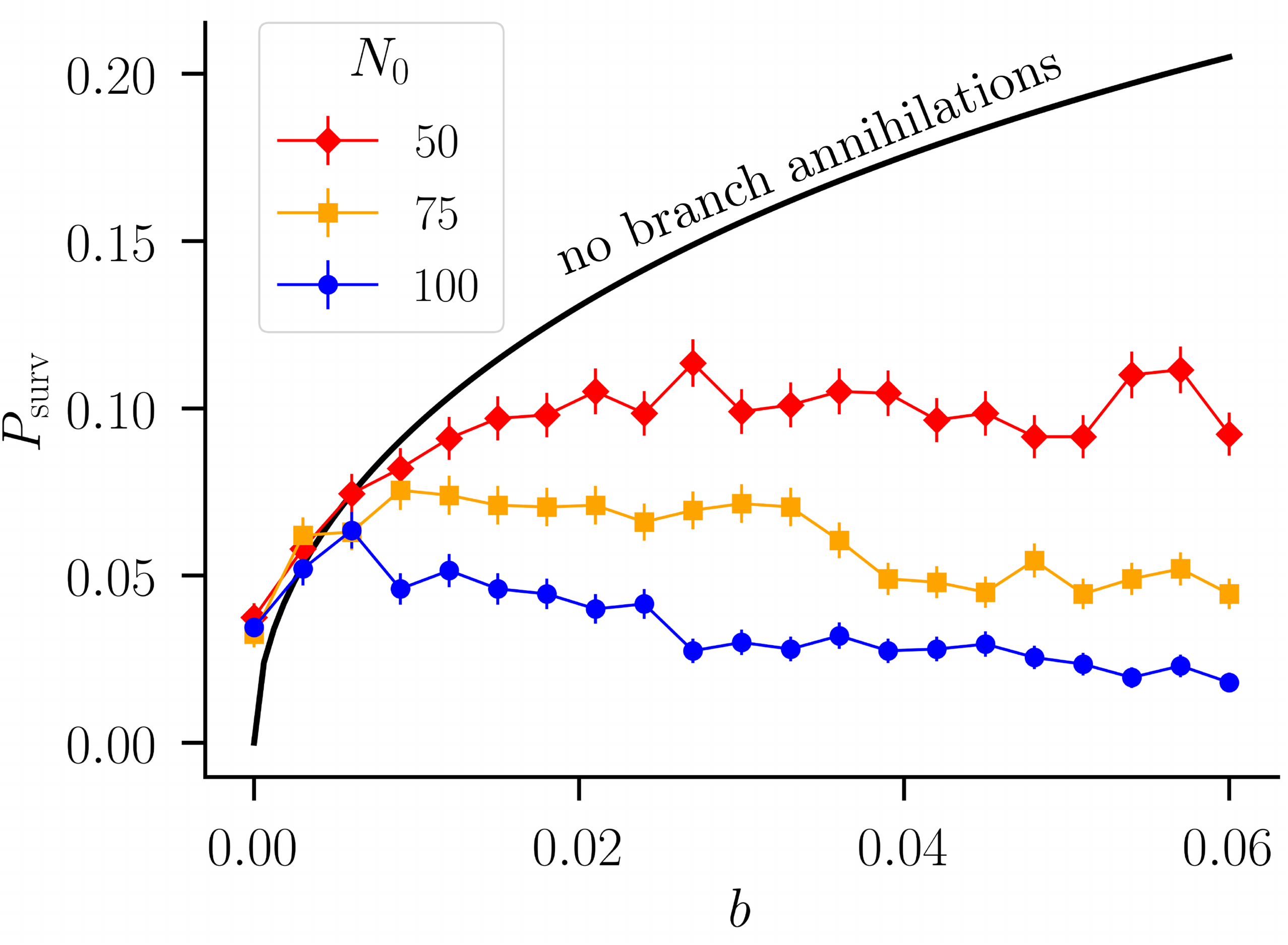}
        \caption{ Survival probability $P_{\mathrm{surv}}$ of a single (neutral) mutant cell on a branched structure after $1000$ generations. Here we include the possibility of branch annihilations but only consider those structures which survive for at least 1000 generations.  The black solid line shows the prediction given by Eq.~\eqref{eq:psurvneutral} for survival on a branching structure without branch annihilation. Note that branch annihilation strongly suppresses the  survival probability for  branching rates $b \gtrsim1/N_0$. This makes intuitive sense as smaller branching rates allow for branches to grow without termination [see Fig.~\ref{fig:branchpics}(c)], while  larger branching rates generate structures with very dense branches which frequently terminate within the population interior [see Fig.~\ref{fig:branchpics}(d)].   \label{fig:BARW_bs}  }
\end{figure}

  On the other hand, for  small selective advantages such that $sN_0 \ll 1 $, inflation will play a significant role as long as the branching rate $b$ is large or comparable to $1/t_e \approx3/(4N_0) $.    
 Thus, for these near neutral mutations, there is a possibility that $b$ is both large enough that inflation has a significant effect \textit{and} small enough that branch terminations do not dominate the survival probability. Our calculations here show that this optimal value of $b$ must be somewhere near  $b \sim 1/N_0$.  Let us now check these arguments in simulations by considering neutral mutations $(s=0)$ specifically.

The results for survival probabilities for neutral mutations at various $N_0$ values are shown in Fig.~\ref{fig:BARW_bs}. We see that the survival probability increases with the branching rate $b$ for small values of $b$, but then starts to decrease  when $b>1/N_0$. This makes sense as we argued previously that the fate of a neutral mutation is determined at long times (proportional to $N_0$) so that branch dynamics should always be relevant in the determination of the survival of a neutral mutation. The inflationary enhancement is most pronounced when the   branching rate is large compared to  $1/t_e$: $b>1/t_e \approx 3/(4N_0)$.  We also showed that  the tips of our branched structures will be more likely to terminate rather than survive when $b>1/N_0$. Thus, there is a potential of a narrow window somewhere around $b \approx 1/N_0$ when branches do not terminate so frequently, but the inflation enhances the survival probability (yielding an ``optimal'' $b$). Our  results in Fig.~\ref{fig:BARW_bs} confirm that, indeed, the survival probability increases with $b$ for all  rates $b \lesssim 1/N_0$. Conversely, when $b \gtrsim 1/N_0$,  the branch annihilations overwhelm the effects of inflation and the survival probability decreases with increasing $b$. Also note that at small $b$, when branch annihilations are less common, the analytic result in Eq.~\eqref{eq:psurvneutral} (black line in Fig.~\ref{fig:BARW_bs}) well approximates the simulation results.

 \begin{figure}[ht]
        \centering
        \includegraphics[width=0.48\textwidth]{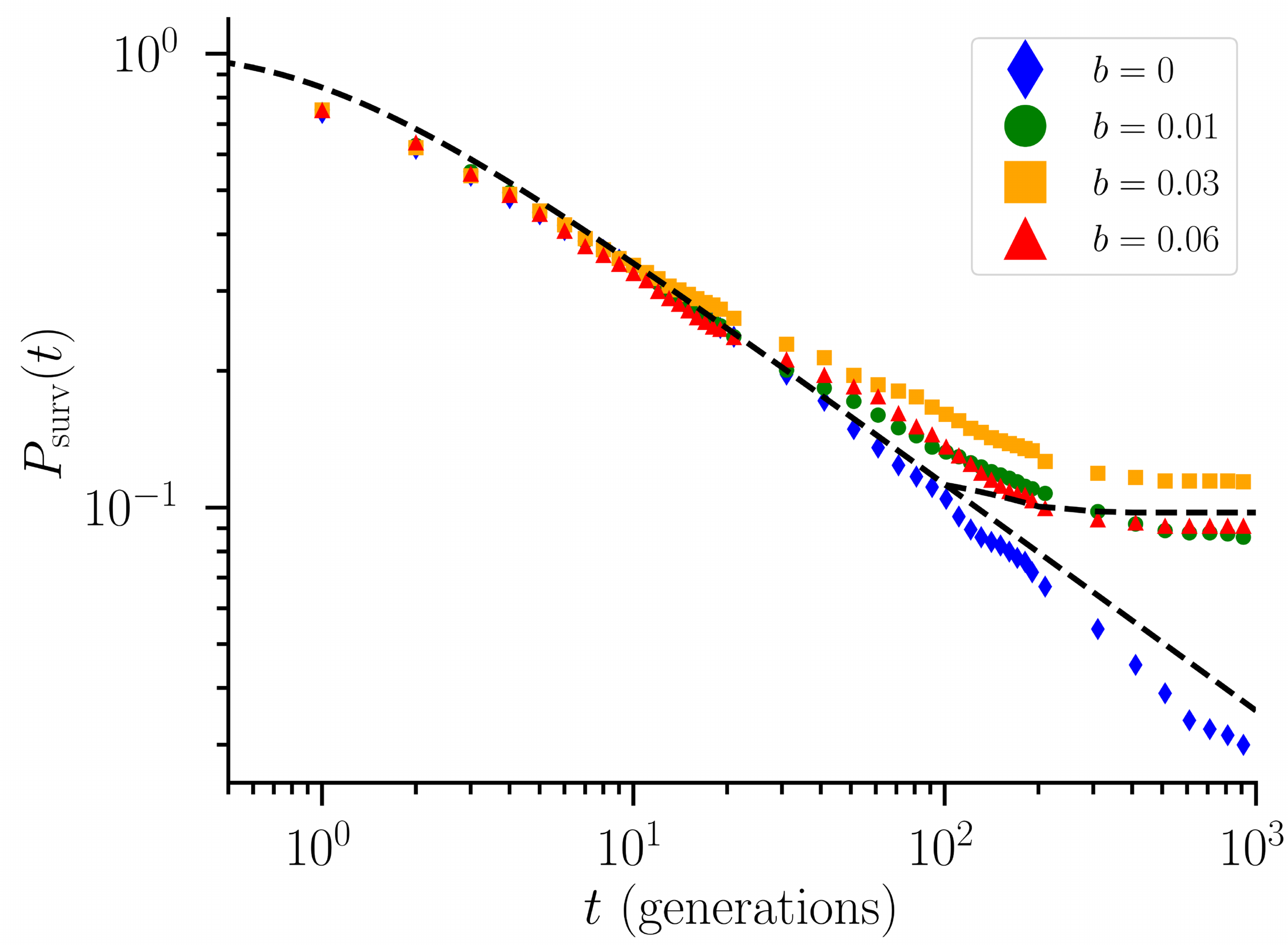}
        \caption{  Survival probability of a neutral mutation as a function of time $t$ in a branching structure with given branching rate $b$ and $N_0=75$ cells (averaged over  2000 branching structures). The points indicate simulation data and the dashed lines are the theoretical results for $b=0$ (lower line) and $b=0.01$ (upper line) from Eq.~\eqref{eq:psurvneutral}. Note that at $b=0$, the probability decays to zero. When $b>0$, the survival probability approaches a limiting non-zero value as $t \rightarrow \infty$. Note that this limiting value is largest for $b=0.03$, and is smaller for both $b=0.01$ and $b=0.06$, indicating that there is an optimal branching rate that enhances the survival.     \label{fig:ProbDecay}  }
\end{figure}

The result in Fig.~\ref{fig:BARW_bs} is interesting as it implies there is an optimal branching rate $b$ for which we get the most  chance for survival of a neutral mutation within the population. In our model, this optimal rate would be approximately $b \approx  1/N_0$. It is also worth emphasizing that a neutral mutation without any branch bifurcation $(b=0)$ would always die out at long times in the large $N_0$ limit. Therefore, although we find that the survival probability decreases with increasing $b$ for $b>1/N_0$, we still expect a non-zero survival probability at long times. In other words, a neutral mutation on a branching and annihilating structure will always have some chance to survive as compared to a neutral mutation in a non-branching population (for large $N_0$). We may see this more vividly by studying the survival probability $P_{\mathrm{surv}}(t)$ as a function of the time $t$.

 The survival probability as a function of time $t$ is shown in Fig.~\ref{fig:ProbDecay} for neutral mutations on branching structures with various $b$ and $N_0=75$ cells. Note that both the theoretical prediction (lower dashed black line) and the simulation results (blue diamonds) show that when there is no branching $(b=0$), the survival probability continues to decay at long times. This is expected as we predict that without branching, the limiting survival probability is  quite small: $1/N_0 \approx0.013$. Conversely, when $b>0$, inflation can \textit{rescue} a neutral mutation and we find that $P_{\mathrm{surv}}(t)$ approaches a limiting value much larger than $1/N_0$ for $t\rightarrow \infty$. We see this in both the theoretical prediction (upper dashed black line) and the simulation data in Fig.~\ref{fig:ProbDecay}. Note that this inflationary effect becomes apparent when $t \gtrsim 1/b$.  This is the expected time at which, on average, a single branch will bifurcate. Finally, note that the survival probabilities in Fig.~\ref{fig:ProbDecay} behave in a non-monotonic fashion with increasing $b$. The $b=0.03$ case gives the largest enhancement relative to $b=0$, but both $b=0.01$ and $b=0.06$ have roughly the same, slightly reduced enhancement. This is another manifestation of the optimal branching rate $b$ that maximizes the survival probability by balancing the effects of inflation and branch termination.

\section{\label{sec:Conclusions}Conclusions and Discussion}

In this study, we modelled cell population survival dynamics on branching geometries, generated by self-avoiding, branching, and annihilating random walks. Our model reproduces a wide range of branching morphologies, ranging from dilute, undulated ducts to highly branching compact tissues. All of these morphologies were achieved via the variation of the actively dividing cell population $N_0$ at the branch tips and the branch bifurcation rate $b$ (in inverse generation times), as shown in Fig.~\ref{fig:branchpics}. We showed that when $1/N_0 <b$,  the branch tips are more likely to annihilate via collisions with existing branches.  This annihilation and the increase in the dividing population size due to branch bifurcation (along with selection) are the   primary driving forces in determining the evolutionary dynamics (i.e., survival) of strains within the population.

Survival probabilities of strains were studied as a function of the branching rate $b$, the mutant selective advantage $s$, and the actively growing population size $N_0$ at each branch tip. 
We find that the branching has a significant effect on survival probability of mutants within the branching population. The branch events themselves serve to increase the effective dividing population and enhance survival probability. Without any branch terminations, the survival probability of a mutation increases monotonically with the branch rate $b$. However, in realistic branching populations, branches will terminate when they become too crowded. In this case, we showed that  as the branched structure grows and branches are culled due to collisions, the survival probability is significantly diminished. We characterize these competing effects and match the simulation results to analytic survival probability calculations. 

We showed that the survival probability of neutral mutations in a branching population with branch terminations is largest at a certain optimal branching rate which balances the effects of inflation and branch annihilation. We also demonstrate that the survival probabilities are largely insensitive to the details of the branch bifurcations and cell growth, being driven by the occurrence of branch collisions and the overall growth in the actively dividing population size. When mutations have a selective advantage $s>0$, we find that the inflation of the population at the bifurcation regions no longer enhances the survival probability (for $sN_0 \gg 1$) as the mutant fate is determined before branch bifurcation events. In this case, it is the branch terminations which govern the survival probability and we find that the probability monotonically decreases with increasing $b$.

There is much opportunity for future work. In this paper we considered just the survival probability of a single mutant occurring in the initial branch of the population. It would be interesting to consider multiple mutations arising at different points in the structure.  One may also track the cell lineages and how they propagate along the branches. This lineage tracking is becoming increasingly possible in real tissues \citep{lungbranch}. We may expect that mutations arising later in the branched tissue development will have a suppressed survival probability if the branched structure is dense with many terminating branch tips. In addition, branched tissues  may have boundary conditions which  stop branch tip growth, such as the fat pads in mammary glands \citep{simons1}.  Cell lineages arising near such boundaries would also have a difficult time propagating. Our model could easily incorporate these features as we keep track of all branch tip populations and would be able to introduce boundaries and additional mutations. Even with these complications, the population evolution would have the same basic features of competition between branch bifurcation and termination in enhancing and suppressing, respectively, the survival probability. 

 To compare with real tissue data, it may also be important to build a more developed cell division model, as real branching tissues may have complex cell rearrangement during the branch growth. Our model captures two major effects on the evolution of such lineages: the increase in the actively-dividing cell population and branch bifurcation events and the extinction of sets of lineages due to branch termination. These effects will be present even in more complicated models of the branch bifurcation and growth. However, we are missing certain aspects such as the feedback of the evolutionary dynamics on the branched structure itself: Selectively advantageous strains should create bulges and deform the branch tips. Finally, it would be interesting to study the effect of changing the branch annihilation radius (the distance between a branch tip and an existing branch at which the branch tip stops growing). In our model, this radius was fixed to be proportional to $N_0$. However, in a realistic branching tissue, a branch tip may stop growing due to some diffusive signal from an existing branch, resulting in a larger annihilation radius.

\section{\label{sec:Acknowledgements}Acknowledgements} 
 
 We thank B. Weinstein and C. Martin for helpful discussions. Computational support was provided by the Advanced Computing Facility at the National Institute for Computational Sciences at the University of Tennessee and Oak Ridge National Laboratory. 
 M. O. L. is grateful for the partial support of the Neutron Sciences Division at Oak Ridge National Laboratory.

\appendix

\section{Model details \label{appx:modeldetails}}

Here we describe some additional details of the branching model. We begin with the evolutionary dynamics as they occur in the actively growing population at each branch tip. The population lives along a ring and in order to efficiently model the competition between cells, we record just the position of the domain wall boundaries for each actively growing population at the branch tip [along with the population size $N(t)]$. Since we   consider one mutant strain,  there are at most two domain walls in each population [see red sector in Fig.~\ref{fig:intro}(a)]. The domain walls are then evolved according to the evolutionary dynamics of the strains (see  \cite{MKNPRE} for more details).   Note that the state of the cells on a given branch has no bearing on the geometry of the branch.

Consider now a branch bifurcation points as shown schematically in Fig.~\ref{fig:intro}(b). At such a point,  a branch is selected to grow according to Eq.~\eqref{eq:growthlaw}. The number of individual cells at the frontier, then, grows according to $\lfloor N(t)\rfloor$, where $\lfloor x \rfloor$ is the greatest integer smaller than or equal to $x$. The time $t$ is measured in generations and increases by 1 for each branch growth step (in which the actively growing population is replaced). Over one generation there are $\lfloor N(t+1)\rfloor-\lfloor N(t) \rfloor$ empty spots that open up in the actively growing population. These empty spots are inserted one at a time between two adjacent cells chosen at random. After insertion, the two chosen cells compete to divide into the newly-opened-up space.  The growth stops when the population grows to a size of  $2N_0$ cells. At this point, the population is divided in half and assigned to two daughter branches.

When the population of a bifurcating ``parent'' branch reaches $2N_0$, the population splits such that each of the daughter branches has an actively growing population of $N_0$ cells.  The split is done by choosing two adjacent cells randomly along the parent branch and assigning each one and the  $N_0-1$   cells to either side of the adjacent cells to form the two daughter branches. The new populations  may contain different portions of the mutant sector (or none at all), depending on the position of the split.  The corresponding domain wall positions for each daughter branch are recorded. A schematic of the process is shown in Fig.~\ref{fig:intro}(b).

\begin{figure}
    \centering
    \includegraphics[height=2in]{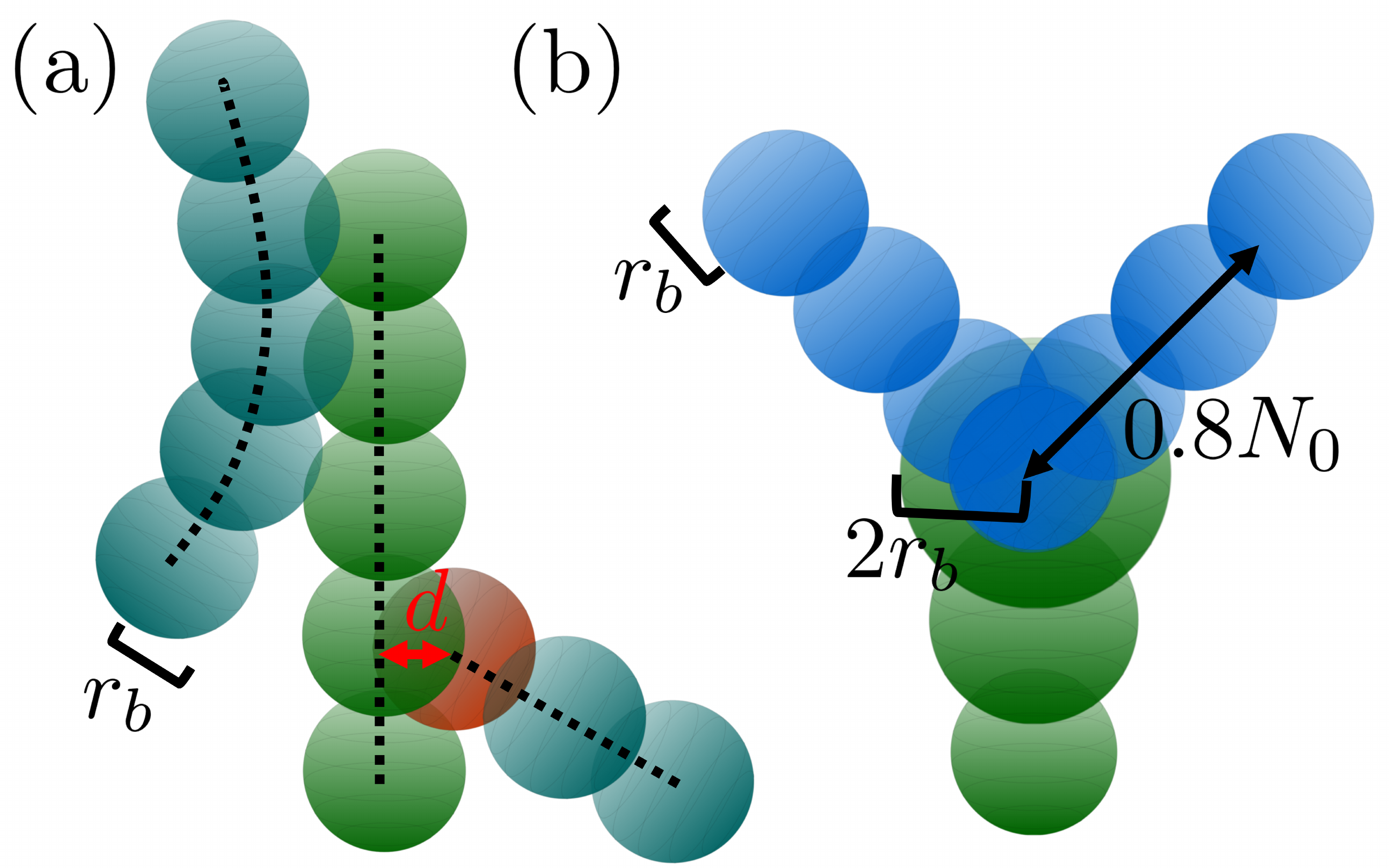}
    \caption{Representations of branch structures with  generations of the actively dividing tip population of size $N_0$  shown as spheres with   radius $r_b=N_0a/2\pi$ equal to the branch cylinder radius. Only a few spheres are shown for clarity.  (a) Three branches in grow in close proximity, with their midlines indicated by dashed lines. The tip of the rightmost branch grows to within a distance $d<r_b$ of the midline of the middle branch, and, thus, terminates, denoted by the red color. The leftmost branch comes within $2r_b$, but not close enough such that the sphere radius overlaps with a branch midline. Growth proceeds in this case. (b) At a bifurcation point, the green parent branch increases in size   [see Eq.~\eqref{eq:growthlaw}] and radius  before splitting into two daughter branches initially occupying the space of the parent branch. The daughter branches, each with radius $r_b$,  grow away from the bifurcation point out to a distance of    $0.8N_0$ before they can undergo annihilation events. }
    \label{fig:branchradii}
\end{figure}

Let us now consider the details of the branch geometry. Each point along the branch is assigned a continuous coordinate and a radius $r_b$ derived from the population size $N_0$ of the branch. A schematic is shown in Fig.~\ref{fig:branchradii}.  Each time a branch grows by one generation at the tip, we check if the branch runs into any other branches that have already been established. A tip will stop growing (``annihilate'' in the language of  random walks) if its radius intersects the midline of any other branch, as shown in Fig.~\ref{fig:branchradii}(a). The ``annihilation distance,'' then, is $d_a=r_b=a N_0/2\pi$ as given in the main text. Also, the self-annihilation rule is modified slightly at a bifurcation point where the actively growing population size increases from $N_0$ to $2N_0$. Here, the daughter branches cannot annihilate due to their proximity to the parent branch until a number of generations greater than $0.8N_0$ after the bifurcation point. This allows for the daughter branches to grow out a little bit before being able to ``collide'' with the parent branch. A schematic of the branching point is shown in Fig.~\ref{fig:branchradii}(b). Moreover, we do not check for self-collisions along a single branch for distances smaller than $0.4 N_0$ between the branch tip and the rest of the branch stalk. This allows for individual branches to grow realistically, with self-collisions only occurring if the tip completely turns around and runs into a previously grown portion of the branch. Given the complex nature of the structures at large branching rates $b$ [see Fig.~\ref{fig:branchpics}(d)], we limited our analysis to branched structures growing for about 1000 generations. Larger times are possible, but computationally intensive. For the purposes of calculating survival probabilities, we average over 2000 instances of the branching structure.

\bibliography{BranchingEvoBib}
\end{document}